%% file: text.tex
\documentclass[prl, preprint, aps]{revtex4-1}
\usepackage{graphicx}
\usepackage{hyperref}

\usepackage{xspace}

\newcommand{\ca}{$\mathrm{^{40}Ca^{+}}$\xspace}
\newcommand{\tran}[2]{#1 $\leftrightarrow$ #2\xspace}
\newcommand{\spt}{\tran{\qs{S}{1/2}}{\qs{P}{1/2}}}
\newcommand{\sdt}{\tran{\qs{S}{1/2}}{\qs{D}{5/2}}}
\newcommand{\dfpt}{\tran{\qs{D}{5/2}}{\qs{P}{3/2}}}
\newcommand{\dtpt}{\tran{\qs{D}{3/2}}{\qs{P}{1/2}}}
\newcommand{\qs}[2]{$\mathrm{#1_{#2}}$}
\newcommand{\mJ}[1]{$m_J=#1/2$}
\newcommand{\Urf}[1]{$U_{\mathrm{rf}}=\SI{#1}{kV}$}
\newcommand{\Udc}[1]{$U_{\mathrm{dc}}=\SI{#1}{V}$}
\newcommand{\apm}[2]{^{+#1}_{-#2}}
\newcommand{\mn}{\ifmmode \langle n \rangle\else $\langle n \rangle$\xspace\fi}
\DeclareRobustCommand{\captex}{\includegraphics{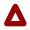} \ce{^{24}Mg^+}, Au
\cite{PhysRevLett.96.253003}; \includegraphics{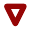} \ce{^{111}Cd^+}, GaAs \cite{Stick2006}; \includegraphics{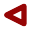} \ce{^{137}Ba^+}, Be-Cu \cite{PhysRevA.65.063407}; \includegraphics{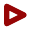} \ce{^{198}Hg^+}, Mo \cite{PhysRevLett.62.403}; \includegraphics{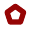} \ce{^{40}Ca^+}, Mo \cite{PhysRevLett.83.4713}; \includegraphics{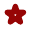} \ce{^{40}Ca^+}, Au \cite{1367-2630-10-4-045007}; \includegraphics{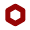} \ce{^{171}Yb^+}, Mo \cite{PhysRevA.61.053405}; \includegraphics{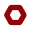} \ce{^{174}Yb^+}, Au \cite{PhysRevA.83.013406}; \includegraphics{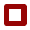} \ce{^{9}Be^+}, Au \cite{PhysRevA.61.063418}; \includegraphics{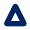} \ce{^{88}Sr^+}, Ag 6K \cite{PhysRevLett.100.013001}; \includegraphics{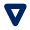} \ce{^{88}Sr^+}, Al 6K \cite{wang:244102}; \includegraphics{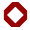} \ce{^{43}Ca^+}, St. Steel \cite{lucas-2007}; \includegraphics{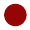} \ce{^{40}Ca^+}, Au} \newcommand{\SI}[2]{\ifmmode #1\enskip\mathrm{#2}\else $#1\enskip\mathrm{#2}$\fi} 
\newcommand{\micro}{\mu} 
\newcommand{\percent}{\%}
\newcommand{\ce}[1]{$\mathrm{#1}$}
\newcommand{\wz}[1]{\ifmmode \nu_z = \SI{#1}{kHz} \else $\nu_z=\SI{#1}{kHz}$}
\newcommand{\wzM}[1]{\ifmmode \nu_z = \SI{#1}{MHz} \else $\nu_z=\SI{#1}{MHz}$}

\begin{document}

\date{\today}
\title{Ground state sideband cooling of an ion in a room temperature trap with
a sub-Hertz heating rate}

\author{G. Poulsen}
\author{Y. Miroshnychenko}
\author{M. Drewsen}
\affiliation{QUANTOP - The Danish National Research Foundation Center for Quantum Optics\\ Department for Physics and Astronomy, Aarhus University, Denmark}

\begin{abstract}
We demonstrate resolved sideband laser cooling of a single \ca\ ion in a macroscopic
linear radio frequency trap with a radial diagonal electrode spacing of 7 mm and
an rf drive frequency of just 3.7 MHz.  For an oscillation frequency of 585 kHz
along the rf-field-free axis, a ground state population of
$99 \pm 1 \percent$ has been achieved, corresponding to a temperature of only
\SI{6}{\micro K}.  For several oscillation frequencies in the range 285 - 585 kHz,
heating rates below one motional quantum per second have been measured at room
temperature. The lowest measured heating power is about an order of magnitude
lower than reported previously in room temperature- as well as cryogenically cooled traps.

\end{abstract}

\maketitle

Resolved sideband laser cooling of trapped atomic ions was originally proposed
with the aim of improving optical spectroscopy \cite{BAPS.20.637}. Since the
first experimental demonstrations of this technique \cite{PhysRevLett.62.403},
single ion cooling to the ground state of the trapping potential has been
performed both in one- \cite{PhysRevLett.62.403} and three-dimensions
\cite{PhysRevLett.75.4011}, with a record high one-dimensional ground state
population of $99.9\%$ \cite{PhysRevLett.83.4713}. The possibility of sideband
cooling several simultaneously trapped ions
\cite{PhysRevLett.81.1525,1464-4266-3-1-357} has furthermore recently led to a
number of outstanding results within quantum information science (See, e.g.,
\cite{Schindler27052011,Ospelkaus2011} and references therein) and ultra-precise
ion spectroscopy \cite{Schmidt29072005,PhysRevLett.104.070802, Chou24092010}. 
While narrow ionic electronic transitions may eventually be applied to establish
new improved optical atomic clocks, the extreme spectral resolution obtainable
with sideband cooled ions enables furthermore fundamental physics
investigations, including search for potential time variation of natural
constants (e.g.,  the fine-structure constant
\cite{PhysRevLett.98.180801,Rosenband28032008} and the proton-to-electro mass
ratio \cite{PhysRevA.71.032505, PhysRevLett.99.150801}), and a potential
electric dipole moment of the electron \cite{PhysRevLett.107.093002}.  For all
the above mentioned applications, uncontrolled interactions between the trapped
ions and the environment are key issues for the final quality of the
measurements. For instance, fluctuating electrical patch potentials on the
trapping electrodes will lead to heating of the ion-motion
\cite{PhysRevA.61.063418}. A simple model of this effect predicts heating rates
which scale roughly inversely with the distance from the ions to the electrodes
to the power of four \cite{PhysRevA.61.063418}, a dependence consistent with
experimental findings \cite{1367-2630-13-1-013032}. For scalable quantum
information processing with trapped ions which requires microscopic traps, the
only viable solution to this problem seems to be using microtraps cooled to
cryogenic temperatures where such fluctuating patch potentials have been found
to be dramatically reduced \cite{PhysRevLett.100.013001}. For spectroscopic
purposes where typically only a single coolant ion and a single spectroscopy ion
are trapped, an alternative strategy is to employ larger macroscopic traps at
room temperature.

In this Letter, we report on one-dimensional ground state sideband laser cooling
of a single \ca ion in a room temperature macroscopic linear radio frequency
trap with a radial diagonal electrode spacing of 7.0 mm and an rf drive
frequency of just 3.7 MHz. In other words, the trap has an ion-electrode spacing
about 3 times larger and an rf frequency nearly a factor of two lower than
reported in previous ground state cooling experiments. In this trap, we have
demonstrated very efficient cooling ($99\pm1\%$ final ground state population)
and measured motional heating rates over months around one motional quantum per
second for oscillation frequencies in the range of 275-585 kHz. These results
strongly indicate that this macroscopic trap at room temperature is very
suitable for high-resolution quantum logic spectroscopy due to the long motional
coherence time and consequently potentially long interaction times
\cite{Schmidt29072005}. The zero point energy associated with the ground state
cooled ion at these low oscillation frequencies ($\sim\SI{6}{\micro K}$ for an
oscillation frequency of \SI{275}{kHz}), make furthermore this trap an
interesting tool for investigations of ultra-cold ion chemistry \cite{B813408C}.
In this connection, it should be mentioned that we have been able to minimize
the amplitudes of the rf sidebands due to uncompensated rf fields to a level
corresponding to a residual micromotion energy equivalent to a temperature below
\SI{1}{\micro K} in all three dimensions.

\begin{figure}
\includegraphics{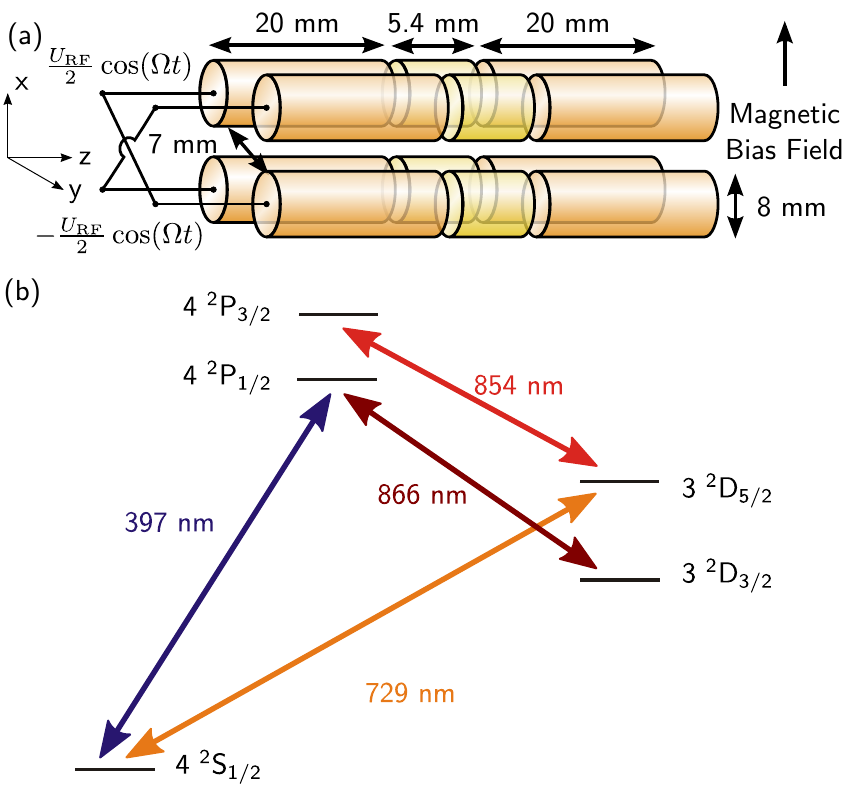}

\caption{(a) Sketch of the macroscopic linear Paul Trap including dimensions of
and electrical potentials on the electrodes. The electrodes colored redish gold
has in addition to the rf potential a dc potential for axial confinement. The
orientation of the applied magnetic bias field is also indicated. (b) Reduced
level scheme for \ca with indication of the transitions addressed under Doppler and sideband cooling. For
details see main text.}\label{fig:trap_geometry}
\end{figure}

In \autoref{fig:trap_geometry}, a sketch of the linear rf trap (details on the
trap design, see Ref. \cite{Drewsen200383}), as well as a reduced level scheme
of the \ca ion with indications of the laser driven transitions are presented.
The linear trap consists of 4 segmented stainless steel cylindrical electrodes
plated with 5 $\mu$m gold to which rf potentials are provided in a symmetrical
fashion to create an effective radial trapping potential. Axial confinement is
achieved through application of an additional dc potential to all eight
end-sections of the electrodes. Smaller dc and rf offset potentials can
furthermore be applied to all twelve sections independently to minimize rf
micromotion at the effective trap potential center. The rf amplitude can reach
1.5 kV and the dc potentials can be raised to 100 V. For typical single ion
experiments, the trap is operated with $U_{\mathrm{RF}} = \SI{1.2}{kV}$ and an
endcap potential of $U_{z} = 20-\SI{80}{V}$ which result in axial and radial
secular oscillation frequencies of \wz{275-585} and $\nu_{x,y} = \SI{1}{MHz}$,
respectively. In all experiments, a bias magnetic field of \SI{40}{Gauss} is
applied along the $x$ direction. Prior to all sideband cooling experiments, the
micromotion is minimized to at least a modulation index below \SI{10^{-1}} in
all directions (along the quadrupole symmetry axis, the $z$-axis, it is even
compensated down to $10^{-3}$) by measuring the Rabi oscillation frequencies on
the micromotion sidebands.

A typical experimental cooling cycle starts with a 5 ms period of Doppler
cooling by addressing the \spt transition of the ion with a 397 nm laser beam
propagating along the $\frac{1}{\sqrt{2}}(\hat{x}+\hat{y})$ direction, and
repumping population out of the \qs{D}{3/2} state by a 866 nm laser beam
propagating along the $\hat{z}$ direction. This is followed by a 8 ms sideband
cooling period comprised of successive pairs of pulses addressing the red
sidebands of the \tran{\qs{S}{1/2}(\mJ{-1})}{\qs{D}{5/2}(\mJ{-5})} transition
and the sideband unresolved \dfpt transition. The cooling sequence starts with
200 cycles on the 2nd red sideband followed by 200 cycles on the 1st red
sideband. Finally, the optical intensity is lowered for another 25 cooling
cycles on the 1st red sideband to minimize the effect of off-resonant scattering
on the carrier transition. During the whole sideband cooling sequence, every 25
cooling cycles are followed by pulses driving the
\tran{\qs{S}{1/2}(\mJ{+1})}{\qs{D}{5/2}(\mJ{-1})} and the \dfpt transitions to
avoid trapping in the \qs{S}{1/2}(\mJ{+1}) state. (The \dfpt and \dtpt
transitions are always driven together to avoid
 optical pumping into the the \qs{D}{3/2} level).

\begin{figure}
\includegraphics{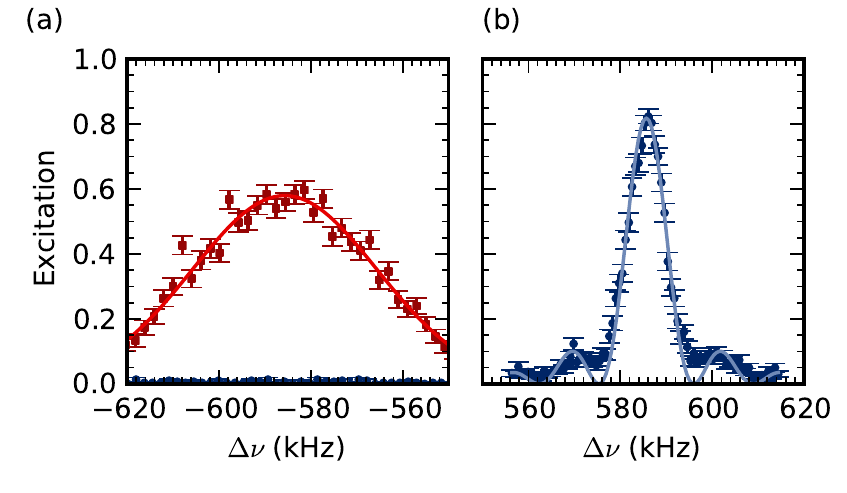}
\caption{(a) First order red sideband excitation spectra before
(red solid squares) and after (blue solid circles) sideband cooling at an axial
oscillation frequency of \SI{585}{kHz}. (b) Corresponding first order blue
sideband excitation spectrum after sideband cooling (blue solid circles).
$\Delta\nu$ is the frequency detuning with respect to the carrier. A quantitative analysis of
the measured strengths indicates a
\SI{0.99\pm0.01} population of the motional ground state. The presented
curves are fits to a Gaussian and Rabi line shape respectively.
}\label{fig:sbc}
\end{figure}

The temperature of the ion is determined by comparing the spectra of the first
red and blue sidebands of the \sdt transition using the standard electron
shelving technique \cite{PhysRevLett.83.4713}. In \autoref{fig:sbc}, examples of
such spectra are presented for an experiment performed with \Urf{1.2} and
\Udc{80}, corresponding to \wz{585}. In \autoref{fig:sbc}a, the red sideband
spectrum is presented both before and after sideband cooling. After cooling,
this sideband clearly vanishes in contrast to the blue sideband presented in
\autoref{fig:sbc}b. A suppressed sideband could in principle also originate from
population in excited motional states with vanishing coupling or pulses
corresponding to $2\pi$ rotations. However, with cooling on both the second and
first sideband (the latter at different powers) and the chosen pulse lengths,
such potential trapping states are avoided.

A quantitative comparison between the two cooled sideband spectra lead
to the conclusion that the ion has an average ground state population of
$0.99\pm0.01$, corresponding to a temperature of \SI{6\apm{1}{6}}{\mu K}. 

\begin{figure}
\includegraphics{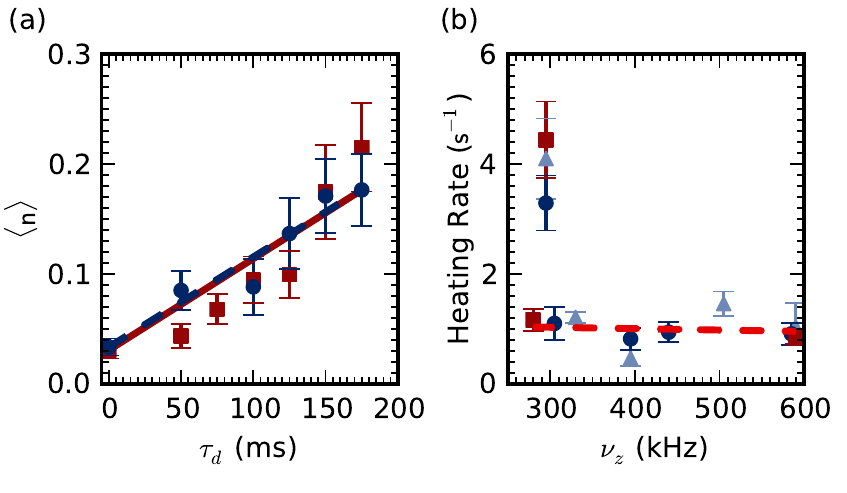}
\caption{(a) The mean occupation number of motional excitation \mn versus time,
$\tau_d$, after ended sideband cooling with an axial oscillation frequency
\wz{585}. The red solid squares (blue solid circles) represent
data point without (with) a mechanical shutter for blocking 397 nm light (see
text for details). The solid red and dashed blue lines are best linear fits to the two
data sets, resulting in heating rates of $d \mn /dt=\SI{0.83 \pm 0.10}{s^{-1}}$ and
$d \mn /dt=\SI{0.84 \pm 0.05}{s^{-1}}$ (b) Measured heating rates versus axial
oscillation frequency. The navy blue circles represent data points obtained the
same day, while the light blue triangular data points are obtained during a a
period of two months. The red squares represents alternative rf voltages to rule out parametric resonances. 
A constant fit to all data points except the ones at
the "resonance" at 295 kHz, gives a heating rate of \SI{1.1 \pm 0.35}{/s}.
}\label{fig:heating_rate}

\end{figure}

To gain information on the reheating of the ion in the trap, the red and blue
sideband spectra have been compared at different delays after cooling from which
the mean occupation number, \mn, has been evaluated. Results from such
measurements are presented in \autoref{fig:heating_rate}a for trapping
conditions identical to those above. As evident from the data, the heating rates
are found to be below one quantum per second. The slight off-set from $\mn=0$ at
short delays arises from non-optimized initial sideband cooling.

To test whether this low heating rate is particular to the chosen axial
oscillation frequency, similar experiments were conducted for other trapping
parameters. In \autoref{fig:heating_rate}b, an extract of these experiments in
terms of measured heating rates as a function of $\nu_z$ in the interval 275-585
kHz is presented. Obviously, the heating rate seems to be essentially
independent on the oscillation frequency and only amounts to one quantum per
second, except for the case of \wz{295}. The low heating rates have been found
to be very persistent, and several of the points in \autoref{fig:heating_rate}b
have indeed been measured over months. This includes the resonance at 295 kHz,
which has also been proven to be independent of the rf voltage applied,
indicating that it is not caused by some particular non-linear resonances due to
an imperfect trap \cite{Alheit1996155}.

\begin{figure}
\includegraphics{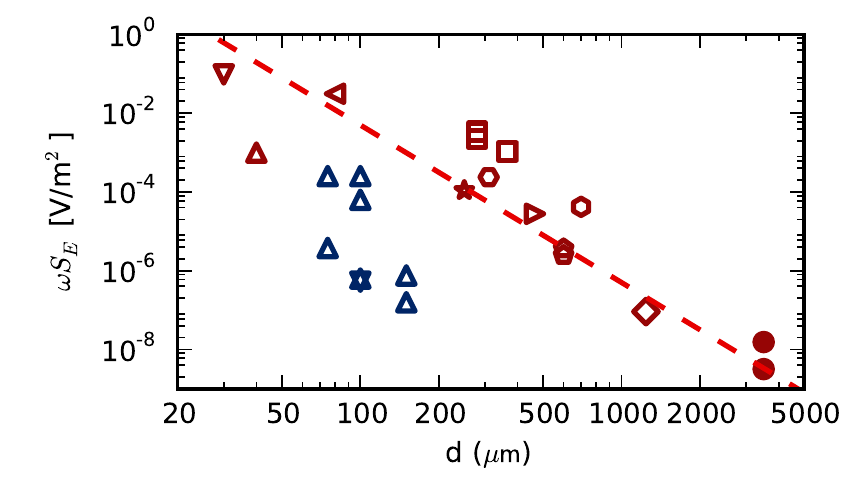}
\caption{Heating rate measurements in terms of the spectral noise density
multiplied by the oscillation frequency ($\omega S_E$) versus the shortest
distance ($d$) between the ion and the trap electrodes. Previous room
temperature (red open data points) and cooled traps (blue open data points)
results are presented together with our results (solid dark red circles).
Obviously, the value of $\omega S_E$ for our experiments with \wz{295} are
nearly two orders of magnitude lower than previously measured, while at the same
time our results generally follow the $1/d^4$ scaling (dashed light red line)
expected for fluctuating patch potentials.
\captex
}\label{fig:spectral_noise}
\end{figure}

In \autoref{fig:spectral_noise}, our heating rate measurements are compared with
ones from other trap experiments on the basis of deduced spectral noise
densities. In terms of the product of the spectral noise density and the
ion oscillation frequency, our results surpasses previous ones obtained in room
temperature as well as cryogenic traps by more than an order of magnitude.

Our measured ultra-low heating rates seem to follow the generally
accepted $1/d^4$ scaling, where $d$ is the nearest distance from the ion to the
electrodes. The dominating contribution is, however, probably from technical
noise and not fluctuating patch potentials, since the measured heating rates are
found to be independent of the oscillation frequency \cite{PhysRevA.61.063418}.

The primary source of motional heating is at present not clear, but a series of
experiments have been conducted to rule out a few. For instance, in
\autoref{fig:heating_rate}a, the two data sets represent two experimental runs
with the only difference being the 397 nm light used for Doppler cooling being
turned off completely by a mechanical shutter after \SI{15}{ms} instead of just
shut of from a level of \SI{45}{mW/cm^2} by an Acousto Optical Modulator
($>\SI{70}{dB}$ extinction) for the one set of data. Apparently, the
non-completely shutting of 397 nm have no significant influence on the heating
rate. Spectrum analysis of the rf and dc source have also been carried out
without any indication of resonances in the
noise spectrum around 295 kHz (within the $\SI{0.5}{dB}$ noise level ).

Though spontaneous emission in connection with sideband cooling at the low trap
frequencies applied in our experiments will lead to heating of the unaddressed
motional degrees of freedom, the low heating rates implies that it should be
feasible to cool all degrees of freedom to the motional ground state (of a
single ion or more) by sequential sideband cooling. Hence, experiments involving
a single sympathetically sideband cooled ion in our type of trap should enable
both high-resolution quantum logic spectroscopy of e.g. highly charged
\cite{PhysRevLett.98.180801} and molecular ions \cite{schmidt:305}, and ion
chemistry in the ultra-cold regime \cite{B813408C}. With respect to the latter
prospect, the low heating rates should additionally make it possible to
adiabatically lower the trap potential to reach temperatures lower than might be
reached by sideband cooling directly.

The large dimensions of our trap furthermore makes it very versatile as it
allows easy introduction of multiple laser beams as well as particle beams
without facing the problem of exposing surfaces close to the ions.

In conclusion, we have demonstrated that it is possible to carry out effective
ground state sideband cooling in a macroscopic linear rf trap with low rf drive
frequencies operated at room temperature. The low heating rates
indicate that such trap could be the proper choice for many
experiments concerning quantum logic spectroscopy and ultra cold chemistry.

\input{text.bbl}

\end{document}

%% file: text.bbl
%